\documentclass[%
 reprint,
 amsmath,amssymb,
 aps,
]{revtex4-2}
\usepackage[dvipsnames]{xcolor}
\usepackage{graphicx}
\usepackage[dvipsnames]{xcolor}
\usepackage{ragged2e}
\usepackage{graphicx}
\usepackage{dcolumn}
\usepackage{bm}
\usepackage{float}
\emergencystretch=\maxdimen
\begin{document}

\preprint{APS/123-QED}

\title{Temporal Walk-off Induced Dissipative Quadratic Solitons}

\author{Arkadev Roy$^1$}
\author{Rajveer Nehra$^1$}%
\author{Saman Jahani$^1$}%
\author{Luis Ledezma$^1$}%
 \author{Carsten Langrock$^2$}
 \author{Martin Fejer$^2$}
 \author{Alireza Marandi$^1$$^*$}
\affiliation{%
$^1$Department of Electrical Engineering, California Institute of Technology, Pasadena, California 91125, USA}%
\affiliation{%
$^2$Edward L. Ginzton Laboratory, Stanford University, Stanford, California, 94305, USA}
\affiliation{$^*$marandi@caltech.edu}

\begin{abstract}
A plethora of applications have recently motivated extensive efforts on the generation of low noise Kerr solitons and coherent frequency combs in various platforms ranging from fiber to whispering gallery and integrated microscale resonators. However, the Kerr (cubic) nonlinearity is inherently weak, and in contrast, strong quadratic nonlinearity in optical resonators is expected to provide an alternative means for soliton formation with promising potential. Here, we demonstrate formation of a dissipative quadratic soliton via non-stationary optical parametric amplification in the presence of significant temporal walk-off between pump and signal leading to half-harmonic generation accompanied by a substantial pulse compression (exceeding a factor of 40) at low pump pulse energies ($\sim$ 4 picojoules). The bright quadratic soliton forms in a low-finesse cavity in both normal and anomalous dispersion regimes, which is in stark contrast with bright Kerr solitons. We present a route to significantly improve the performance of the demonstrated quadratic soliton when extended to an integrated nonlinear platform to realize highly-efficient extreme pulse compression leading to formation of few-cycle soliton pulses starting from ultra-low energy picosecond scale pump pulses that are widely tunable from ultra-violet to mid-infrared spectral regimes.  
\end{abstract}  

\maketitle

Formation of dissipative solitons in nonlinear resonators has become a versatile mechanism for stable femtosecond sources \cite{kippenberg2018dissipative, herr2014temporal}. In the frequency domain it corresponds to a broadband frequency comb which, when self-referenced, leads to a myriad of applications in precision measurements spanning from spectroscopy \cite{dutt2018chip, muraviev2018massively}, astro-combs \cite{suh2019searching,obrzud2019microphotonic}, atomic clocks \cite{stern2020direct}, ranging \cite{suh2018soliton,trocha2018ultrafast}, and imaging \cite{ji2019chip}, to name a few. Recently, the ambit of frequency combs has expanded to cover promising avenues including massively parallel data communication \cite{marin2017microresonator}, and realization of machine learning accelerators \cite{ xu202111}. To cater to this increasing list of technologically important applications, there lies the outstanding challenges of attaining low-power operation \cite{Lu:21}, high pump to soliton conversion efficiency \cite{xue2019super, xue2017microresonator, bao2014nonlinear, obrzud2017temporal, bao2019laser}, broadband (octave-spanning and widely tunable) comb formation in a compact platform \cite{stern2018battery, shen2020integrated}, reliable fabrication and operation of high-Q resonators which need to be addressed. 

In the past decade there has been extensive research on Kerr based frequency combs where a $\chi^{(3)}$ nonlinear resonator is coherently driven by a continuous wave (CW) laser to excite temporal solitons \cite{herr2014temporal}. However, the Kerr nonlinearity, being a cubic nonlinearity, is inherently weak, and so requires the use of high Q resonators to reach the threshold of parametric oscillation with reasonable pump power. The frequency comb forms around the driving CW laser (which itself typically constitutes a comb line), and it requires precise dispersion engineering and in some cases,  use of multiple pump lasers \cite{moille2021ultra} to extend the comb to its harmonics and sub-harmonics. These issues can be alleviated by operating with quadratic nonlinearity which can typically cause significant nonlinear mixing at power levels that are orders of magnitude lower than its cubic counterpart \cite{bruch2020pockels}. With the ability to perform harmonic conversion through properly phase-matched quadratic nonlinear interactions, $\chi^{(2)}$ nonlinear media promise an ideal platform to realize widely tunable self-referenced frequency combs. Although frequency comb generation through quadratic nonlinearity \cite{mosca2016direct, mosca2018modulation, parra2019frequency, leo2016walk, nie2021deterministic, nie2020quadratic} and femto-second optical parametric oscillators  \cite{leindecker2012octave} have been the subject of several theoretical and experimental investigations, demonstration of quadratic soliton formation remains sparse \cite{bruch2020pockels, jankowski2018temporal}.   \\

Fundamental limits on the efficiency (pump to soliton conversion) of CW-pumped Kerr solitons \cite{bao2014nonlinear, xue2017microresonator} have motivated study of their pulsed-pump driven arrangements \cite{obrzud2017temporal, obrzud2019microphotonic}. Modulated pumps also provide additional control on the dynamics of the temporal solitons \cite{cole2018kerr, jang2015temporal, erkintalo2021phase}. Additionally, synchronous pumping allows direct control on the repetition rate of the ensuing driven soliton.  Another route to achieve high conversion efficiency is to use low finesse cavities with large outcoupling, where the round-trip loss is compensated by proportionate amplification \cite{englebert2021temporal}. Thus soliton formation in a synchronously pumped low-finesse quadratic nonlinear resonator represents a viable route to realize highly efficient widely tunable broadband frequency combs \cite{jankowski2018temporal}.  \\   

In this work,  we demonstrate walk-off induced temporal solitons in a degenerate optical parametric oscillator (OPO) based on pure quadratic nonlinearity. We show that the quadratic soliton can be supported in both normal and anomalous group-velocity dispersion regimes. We also show that this quadratic soliton exists in a low-finesse optical cavity which can lead to high conversion efficiency. We achieved giant pulse-compression exceeding a factor of 40 at picojoule level pump energy. We investigate the dynamics of this quadratic soliton and characterize its different regimes of operation. Furthermore, we define a figure of merit which can act as the design guideline for achieving extreme pulse compression and optimum soliton formation in a dispersion engineered cavity that can be accessible through integrated platforms. Our results pave the way for the generation of energy-efficient dissipative quadratic solitons breaking some of the barriers for generation of Kerr solitons which demand high Q cavities, feature limited conversion efficiency, require anomalous dispersion for bright soliton formation, and possess limited wavelength tunability. \\

\begin{figure}
\centering
\includegraphics[width=0.45\textwidth]{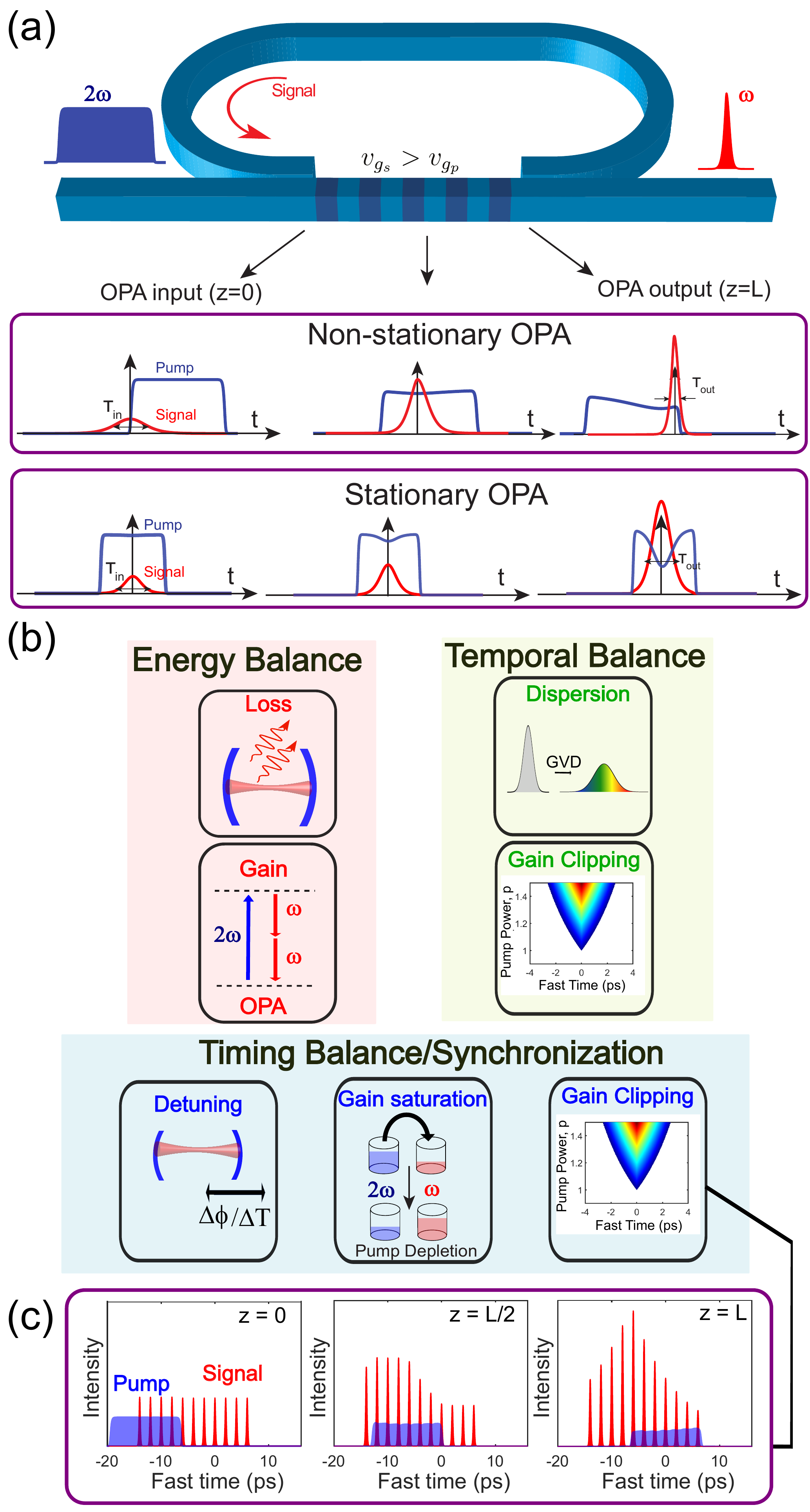}
\caption{\label{fig: schematic} \textbf{Walk-off induced quadratic soliton formation process }. a) Schematic of the doubly resonant half-harmonic synchronously pumped OPO with resonant signal and non-resonant pump. The quadratic nonlinear interaction happens in the periodically poled region and, owing to the large GVM between the signal and the pump, significant pulse compression occurs due to the non-stationary OPA process as the signal walks off through the pump. This is contrasted with the stationary OPA case which features negligible temporal walk-off, where the signal amplification is not accompanied by considerable pulse compression. b) The dissipative soliton is sustained in the OPO by a triple balance of energy, temporal broadening, and timing. The energy loss through dissipation is balanced by the parametric gain through the OPA process, while the temporal broadening due the group-velocity dispersion is arrested by the temporal gain-clipping mechanism. Finally, the timing synchronization is achieved by the delicate balance between linear cavity detuning, nonlinear acceleration due to gain saturation, and gain-clipping mechanism. c) Illustration of the temporal gain-gating mechanism, i.e., the dependence of gain on the relative delay between the pump and signal pulses, which is responsible for the gain-clipping. The signal pulse that experiences maximum temporal overlap with the pump pulse throughout the non-stationary OPA process extracts the highest gain.}
  \end{figure}

\section{Results}

We consider a degenerate OPO \cite{hamerly2016reduced} as illustrated in Fig. 1(a) (see Supplementary Section 2 for detailed schematic). The OPO is driven synchronously by a pump pulse with a temporal width of several picoseconds. The quadratic nonlinear interaction takes place in a periodically poled lithium niobate waveguide \cite{langrock2007fiber, marandi2015guided} providing parametric interactions between the pump at the fundamental frequency and the signal at the half-harmonic frequency. The cavity is completed with a combination of polarization maintaining fibers and a suitable free-space section to ensure that the pump repetition rate is approximately equal to multiples of the cavity free-spectral range. Different lengths of anomalous and normal dispersion (dispersion shifted) fiber combinations are used to manipulate the cavity dispersion.  \\

The walk-off induced quadratic soliton formation is supported by a non-stationary optical parametric amplification  (OPA) process \cite{akhmanov1968nonstationary, khaydarov1995pulse}. The temporal soliton at the half-harmonic (1550 nm) walks through the pump (775 nm) due to the group velocity mismatch (GVM). This non-stationary OPA process results in nonlinear effects at a given point and time in the waveguide that depends upon the field magnitude at other times (see Supplementary Section 3). This allows a signal pulse that is much shorter than the pump pulse  to extract most of the pump energy. This non-stationary OPA leads to significant pulse compression alongside signal amplification. In contrast, in a stationary OPA process (with negligible temporal walk-off between the pump and signal), the signal gets amplified without considerable pulse compression as shown in Fig. 1(a) \cite{ledezma2021intense}. Figure 1(b) shows the mechanisms contributing to the nonlinear dynamics of the OPO which are responsible for this quadratic soliton formation. The dissipative soliton loses energy via different dissipation pathways of the low-finesse cavity which include intrinsic cavity round-trip loss and the out-coupling (ideally the cavity should operate in the highly over-coupled scenario). This energy loss is counterbalanced by the photon addition due to the parametric amplification process, where the soliton gains energy from the pump pulse. The OPO oscillation threshold occurs when the parametric gain overcomes the loss, and the quadratic soliton is supported near threshold.

 \begin{figure*}
\centering
\includegraphics[width=1\textwidth]{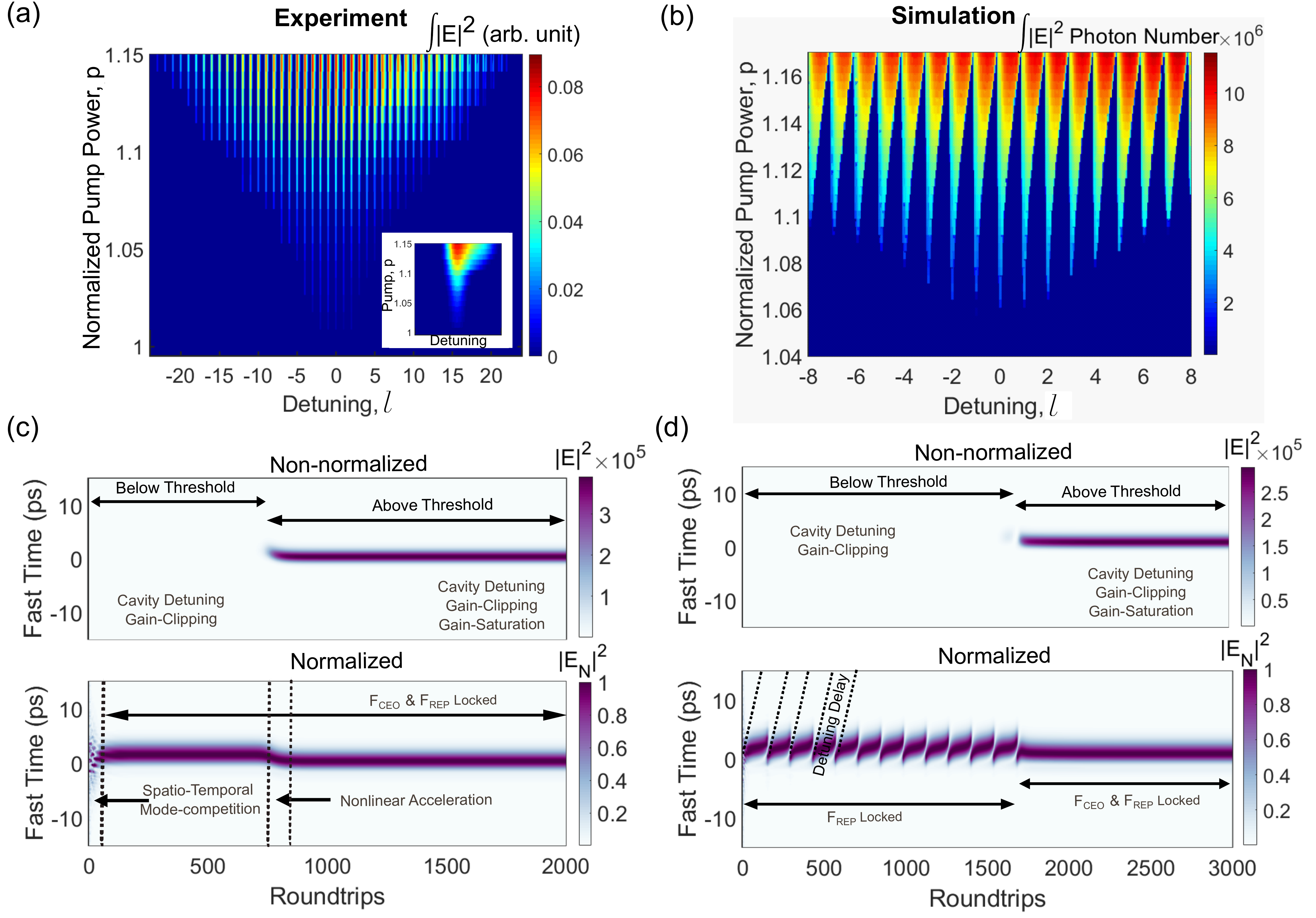}
\caption{\label{fig: 6} \textbf{Cavity detuning dependence of doubly resonant OPO and its impact on synchronization. } Mode-locking range of doubly resonant sync pumped OPO as a function of cavity detuning obtained a) experimentally and b) via numerical simulation. OPO oscillation occurs around discrete peaks centered at integer values of detuning parameter $l$. The zoomed-in view of a single peak is shown in the inset. c) Evolution of the intra-cavity OPO field from random noise to the sech shaped signal soliton pulses in the steady state.(top panel: Non-normalized intensity; bottom panel: Normalized intensity to highlight different dynamical regimes). d) Evolution of the intra-cavity OPO field in the case of large cavity detuning ($l = 50$) highlighting the importance of gain-saturation and gain-clipping in timing balance/ synchronization (top panel: Non-normalized intensity; bottom panel: Normalized intensity to highlight different synchronization regimes in large detuning scenario).}
\end{figure*}

 The group-velocity dispersion (GVD) of the medium (waveguide + cavity) leads to temporal pulse broadening. This GVD-induced broadening is prevented by the temporal gain-clipping mechanism (see Supplementary Section 3) \cite{hamerly2016reduced}. The pulsed pumping scheme in the synchronously driven OPO leads to a temporal gain window which is responsible for the time gating of the parametric gain and is expressed by the gain-clipping effect. In Fig. 1(c) we consider the unsaturated amplification of several signal pulses with different temporal positions on the fast time scale with respect to the pump pulse. The signal pulse which undergoes the maximal overlap with the pump in the entire non-stationary OPA process experiences the maximum gain. This gain gradually decreases on either side of this optimal temporal position on the fast-time axis, thereby enforcing a temporal gating of the gain. This gain window progressively broadens as the pump power is increased above threshold. A signal pulse which experiences GVD-induced temporal broadening and extends beyond this temporal gain window will experience less net gain. This competition between GVD and this gain-clipping mechanism gives rise to the temporal balance. \\
 
 Finally, the timing balance (synchronization) is determined by the mutual interplay between the linear cavity detuning, gain saturation, and the gain clipping. The cavity detuning causes a timing mismatch from exact synchrony with the pump repetition rate. Doubly resonant OPOs can only oscillate around cavity detunings where the round-trip phase accumulation are integer multiples of $\pi$ \cite{hamerly2016reduced}. The cavity detuning phase can be expressed as $\Delta \phi = \pi l$, where integer values of $l$ represents the center of these discrete OPO peaks as shown in Fig. 2(a,b). With increasing pump power, more peaks (larger values of $|l|$) go above their respective oscillation thresholds as shown in Fig. 2(a,b). A zoomed-in view of a single peak structure is shown in the inset of Fig. 2(a). Within a peak, the OPO can oscillate in degenerate or non-degenerate regimes \cite{roy2021spectral}. We operate at the cavity detuning which corresponds to the degenerate mode of operation in order to access the dissipative quadratic soliton. The timing mismatch (detuning induced delay with respect to the synchronous pumping) can be expressed as $\Delta T= \frac{\lambda l}{2c}$, where $c$ is the group velocity of the half-harmonic signal with wavelength $\lambda$ in the cavity.  \\
  
\begin{figure*}
\centering
\includegraphics[width=1\textwidth]{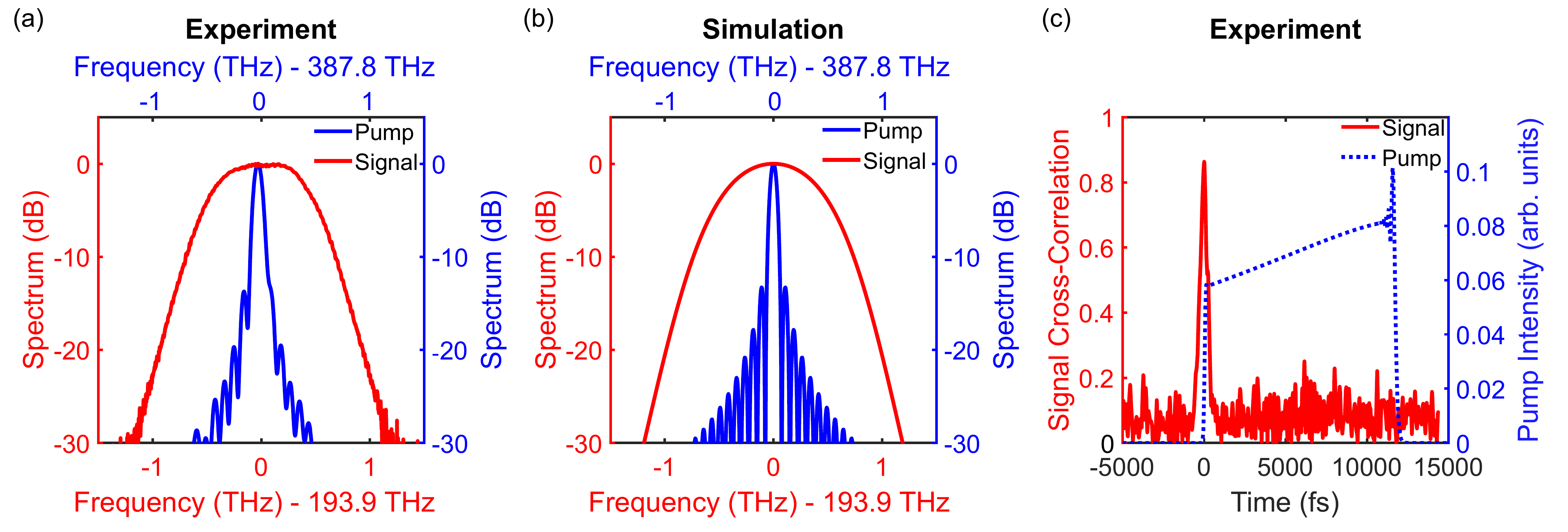}
\caption{\label{fig: Fig2} \textbf{Spectral and temporal characteristics of the quadratic soliton}. Spectrum of the soliton (signal at half-harmonic centered around 193.9 THz represented by red solid line) obtained a) experimentally and b) through numerical simulation. The pump spectrum is represented in blue solid lines centered around 387.8 THz. c) The degenerate half-harmonic quadratic soliton formation is accompanied with significant temporal pulse-compression as can be visualized from the intensity cross-correlation trace. The flat-top pump temporal profile ($\sim 13.2 \textrm{ps}$) is shown as the blue dotted line and is obtained by numerical simulation.   }
\end{figure*}

 The gain saturation arises due to the pump depletion. As the signal walks through the pump (due to the group velocity mismatch) the leading edge of the soliton experiences larger gain than the trailing edge, since the trailing edge experiences the depleted pump. This causes a pulse centroid shift in response to this underlying nonlinear acceleration \cite{jankowski2018temporal}. The gain-clipping on the other hand is aligned with the pump temporal position on the fast-time scale and also contributes to the timing balance. The synchronization of the quadratic soliton supported in the synchronously pumped OPO can be expressed in terms of $F_{CEO}$ (Carrier Envelope Offset Frequency) locking, and $F_{REP}$ (Repetition Frequency) locking to the pump. While the doubly resonant OPO can have multiple modes of operation (see Supplementary Section 8), the quadratic soliton in the steady state exists in the $F_{REP}$ and $F_{CEO}$ locked synchronized state. We note that the output of an OPO oscillating at degeneracy is phase-locked to its pump, in stark contrast with doubly-resonant non-degenerate and singly resonant cases \cite{roy2021spectral}. Thus the demonstrated quadratic soliton is capable of transferring the frequency comb stability properties of the pump from near-infrared to mid-infrared spectral regions in addition to the coherent spectral broadening (i.e. the pulse shortening) mechanism. The evolution of these solitonic pulses from background noise obtained via numerical simulation is shown in Fig. 2(c). The non-normalized version (top panel of Fig. (2c)) and normalized version (bottom panel of Fig. 2(c), where the intra-cavity intensity of each round-trip is normalized to itself) of the round-trip evolution highlights the several dynamical regimes of the quadratic soliton formation. The nonlinear dynamics is initiated by spatio-temporal gain competition, followed by gradual building up of the signal/soliton pulse which eventually becomes intense enough to undergo gain-saturation leading to pulse centroid shift/nonlinear acceleration. Finally, the soliton reaches the steady-state and maintains synchronization with the pump. Figure 2(d) shows the time evolution of the OPO in the case of large cavity detuning. This exemplifies the role of gain-saturation and gain-clipping in timing balance/synchronization. Large cavity detuning is also associated with timing mismatch (from exact synchrony) which results in a pulse delay/ advance each round-trip, as shown in the initial evolution cycles in bottom panel of Fig. 2(d). The slope of this pulse delay is represented by the dotted lines and corresponds to the cavity detuning-induced timing delay ($\Delta T$). Below threshold, this large timing mismatch cannot be completely compensated by the gain-clipping mechanism leading to an  $F_{CEO}$ unlocked state, where the signal still maintains $F_{REP}$ locking. Above threshold, onset of gain-saturation takes place, and the combined effect of gain-saturation and gain-clipping leads to the timing balance resulting in the soliton maintaining synchronization ($F_{CEO}$ and $F_{REP}$ locked ) with the pump. The combination of these balancing effects can be elucidated using the semi-analytical variational formalism which expresses the pulse parameters (energy, temporal width, and centroid) in terms of the cavity and driving parameters (see Supplementary Section 3.2) \cite{hamerly2016reduced}.   \\

\begin{figure*}
\centering
\includegraphics[width=1\textwidth]{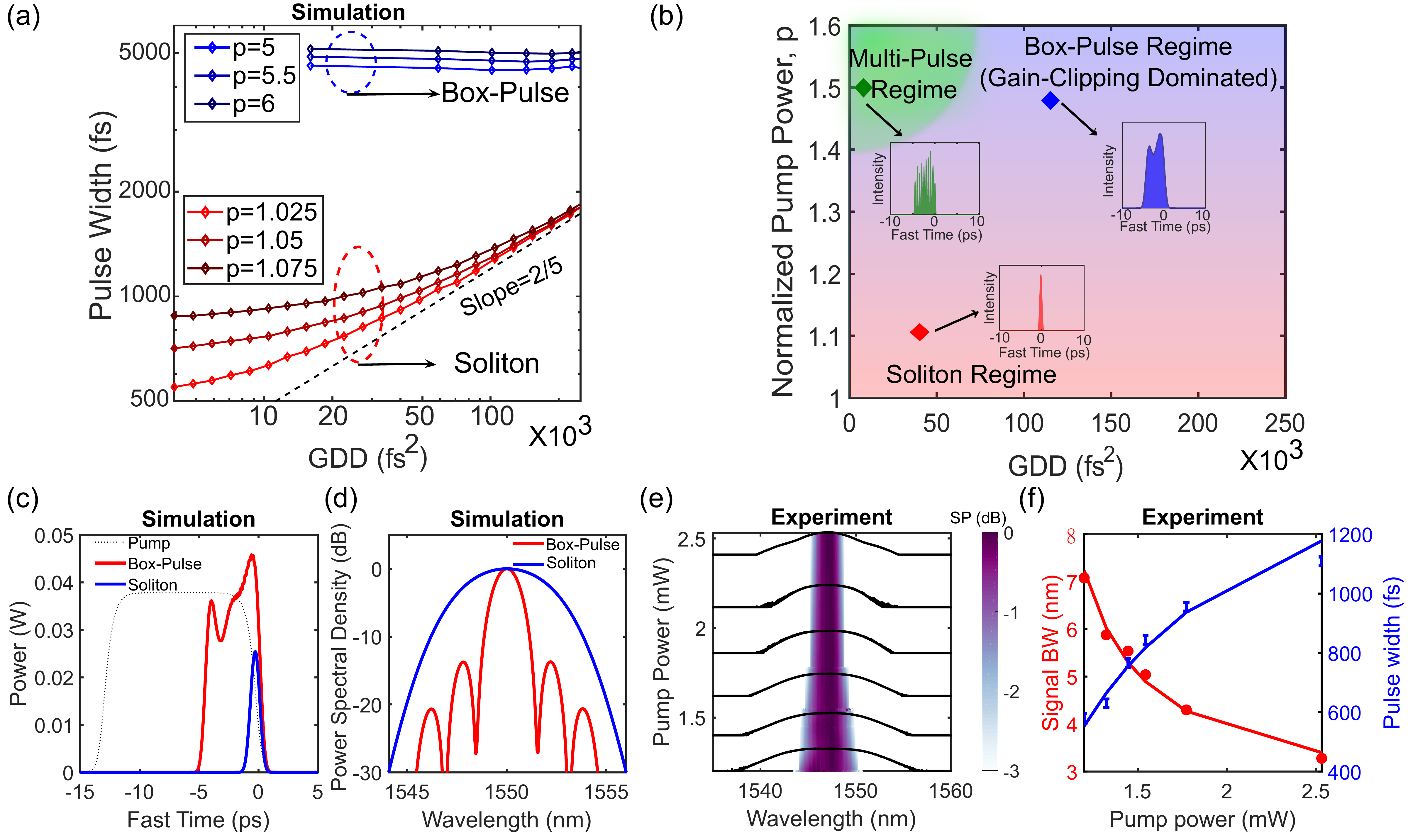}
\justifying
\caption{\label{fig: Fig3}  \textbf{Soliton and Box-pulse regimes.}  a) Pulse width scaling as a function of dispersion in the soliton and box-pulse regimes. The soliton pulse width scales linearly (in log-scale) with dispersion, while the pulse width in the box-pulse regime, which is dominated by gain-clipping, is almost independent of dispersion. b) Illustrative phase diagram in the parameter space of pump power and dispersion indicating the region of existence of various pulse regimes (soliton, box-pulse, multi-pulse). c) Temporal profiles and d) spectrum of the OPO in the soliton and box-pulse regimes respectively. The doubly resonant OPO supports quadratic soliton near threshold, and it transits to box-pulse shaped pulses as the pump power is raised above threshold. e) Spectral narrowing as the OPO enters further in the box-pulse regime as is evident by the decreasing OPO 3 dB bandwidth. The shading represents the 3dB portion of the spectra. f) The associated temporal broadening reflected in terms of temporal full-width half maximum obtained from the intensity cross-correlation data as the pump power is increased above threshold. The solid curves represents fit according to the gain-clipping variation. The error bars represent uncertainty associated with the intensity cross-correlation technique. }
\end{figure*} 

 Figure 3 shows the measured and simulated spectral and temporal characteristics of the quadratic soliton. Significant spectral broadening of the signal compared to the pump is shown in Fig. 3(a,b). The soliton pulse is characterized using an intensity cross-correlation technique as shown in Fig. 3(c) overlaid with the pump pulses at the output of the waveguide. The OPO operation at degeneracy is confirmed by the radio-frequency (RF) beat-note measurement (see Supplementary Section 5).

 The quadratic soliton is formed near the oscillation threshold of the OPO. As the pump power (expressed in normalized form $p$ denoting the number of times above threshold) increases further above threshold, the temporal width of the gain-clipping region increases, and the soliton transitions into the box-pulse regime (see Supplementary Section 4) \cite{hamerly2016reduced}. In the box-pulse regime, gain-clipping dominates over the cavity GVD, and the pulse assumes a box-pulse shape, deviating from the sech profile in the near-threshold soliton regime. In the soliton regime, the effect of group-velocity dispersion is counterbalanced by the gain-clipping, and the soliton exhibits characteristic pulse-width variation with a power law dependence on the total cavity dispersion (GDD, group delay dispersion) (scale to the 2/5 power of the GDD, see Supplementary Section 3.2), while in the box pulse regime, the pulse length is almost invariant with GDD. This distinct pulse-width scaling is shown in Fig. 4(a). Figure 4(b) shows an illustrative diagram of the region of existence of different pulse regimes in the parameter space of pump power and dispersion. The soliton regime is accessed close to threshold. When the pump power is far above threshold, the system enters the box-pulse regime for large values of dispersion, while multi-pulsing occurs for small values of dispersion. This trend follows both in normal and anomalous dispersion regimes. Typical spectral and temporal characteristics of the OPO in the solitonic and box-pulse regime are shown in Fig. 4(c,d). The 3 dB spectral bandwidth of the OPO decreases with increasing pump power as shown in Fig. 4(e), and the corresponding variation of the full-width at half-maximum of the pulses in time domain as measured using the intensity cross-correlation technique is shown in Fig 4(f). This scaling of the pulse width is distinct from the temporal simultons, which are bright/dark soliton pairs in a quadratic media (see Supplementary Section 11) \cite{trillo1996bright}. In an OPO operating in the simulton regime, the spectral bandwidth increases with increasing pump power \cite{jankowski2018temporal}.\\   
 
\begin{figure*}[!ht]
\centering
\includegraphics[width=1\textwidth]{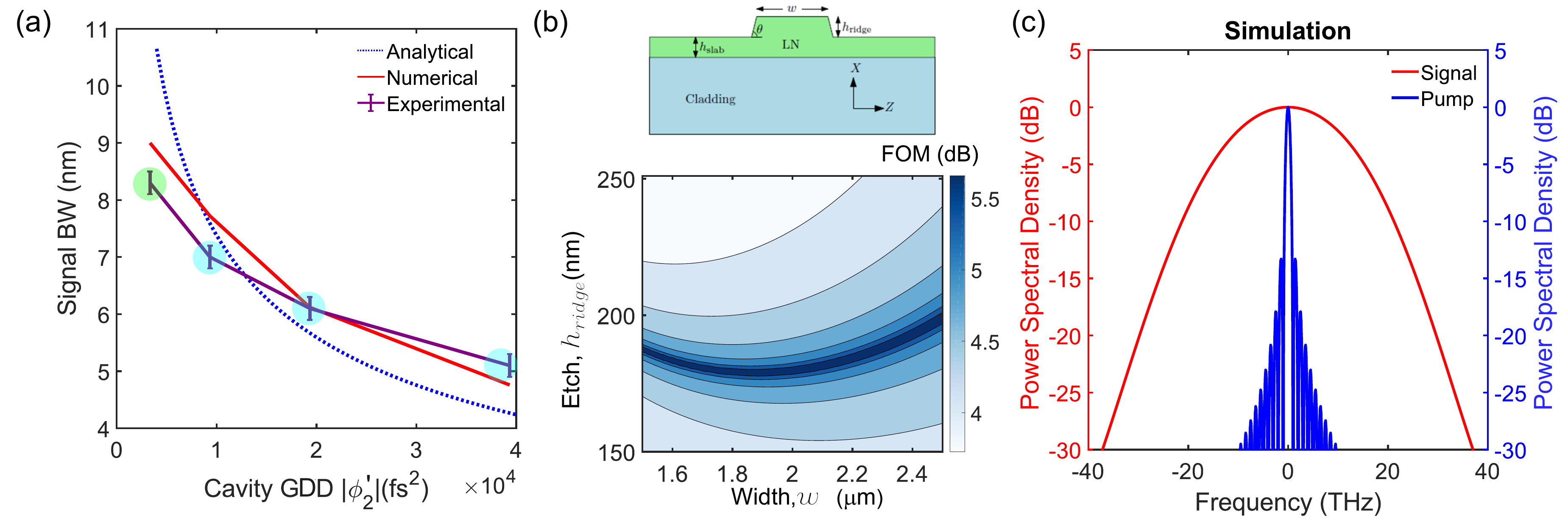}
\justifying
\caption{\label{fig: Fig4} \textbf{Dispersion engineering and efficient half-harmonic soliton pulse-compression.} a) Variation of the soliton bandwidth as a function of cavity dispersion. Different dispersion configurations are realized in the experiment by altering the combinations of normal (blue circles) and anomalous dispersion (green circles) fibers. The error bars represent variations due to the uncertainty in the number of times above threshold operation in different GDD scenarios. b) Fine control in the dispersion engineering is possible in thin-film integrated waveguide based devices by choosing appropriate etch depth and waveguide width. The FOM defined in (Eq. 1) is plotted, where high values of FOM indicate large attainable pulse compression. Here, $L$ is assumed to be 6 mm.  c) Numerical simulation corresponding to the optimum FOM showing significant spectral broadening in the soliton formation, corresponding to a pulse-compression by a factor of $\sim$ 60. }
\end{figure*}  

The soliton pulse width can be obtained from the semi-analytical variational calculations (see Supplementary Section 3), and its dependence on the cavity parameters can be expressed as: $\tau_{sech}= \left(\frac{7}{15}\frac{\left(\phi_{2}^{'}\right)^{2}T_{p}}{\textrm{ln}(G_{0})\textrm{ln}(2)}\right)^{\frac{1}{5}}$, where $\phi_{2}^{'}$ is the cavity GDD, $(1-G_{0}^{-1})$ represents the round-trip loss, and $T_{p}$ is the pump pulse width. We have assumed the optimum pump pulse width ($T_{p}=Lu$, where $L$ is the waveguide length and $u$ is the walk-off parameter) where the pulse width matches the walk-off length in the waveguide (see Supplemental Section 9). Also, the contribution of higher-order dispersions has been neglected in deriving this expression. In the limit of GDD $\rightarrow$ 0, the above GDD dependent approximation of the pulse width breaks down, and the pulse width is instead determined by the higher-order dispersion coefficients (see Supplemental Section 3.3). The pulse compression factor can then be expressed as $T_{p}/\tau_{sech}$. We define a figure of merit (FOM) which is indicative of the amount of pulse compression attained and can serve as the design guidelines for achieving optimum soliton compression. The FOM for a given length ($L$) of the phase-matched quadratically nonlinear region is defined as (see Supplemental Section 3):\\

\begin{equation}
 FOM= \left |\frac{u^{2}L}{\beta_{2}} \right |
\end{equation} \\

A large value of GVM and small value of GVD ($\beta_{2}$) (either normal or anomalous) favours efficient soliton compression. Figure 5(a) shows the experimental results of variation of soliton spectral bandwidth with changing cavity dispersion that has been realized by combining different lengths of normal and anomalous GVD fibers. Some of these fiber combinations yield net normal GVD, and the rest experiences net anomalous GVD. The quadratic soliton can exist irrespective of the sign of the cavity second order GVD coefficient. Extensive dispersion engineering capability can be accessed through integrated nanophotonics platform which can be designed to maximize the FOM \cite{ledezma2021intense, jankowski2020ultrabroadband, jankowski2021efficient}. Fig. 5(b) shows the plot of the FOM as the width and etch depth of a typical lithium niobate ridge waveguide is varied (see Supplemental Section 6). If we now consider a point corresponding to a large value of the FOM, we predict (using numerical simulation) a pulse compression in excess of a factor of 60 for a 6-mm-long parametric gain section, leading to the generation of few optical cycles pulses starting from picosecond pump pulses as shown in Fig. 5(c). Further improvement can be obtained by utilizing longer gain sections and engineering a flat dispersion profile while taking higher-order GVD coefficients into consideration (see Supplemental Section 3.3) \cite{dietrich2020higher}. In the limit where the entire resonator provides parametric gain, it maybe possible to generate highly efficient CW-driven walk-off induced quadratic soliton.  \\  

\begin{figure}[!ht]
\centering
\includegraphics[width=0.5\textwidth]{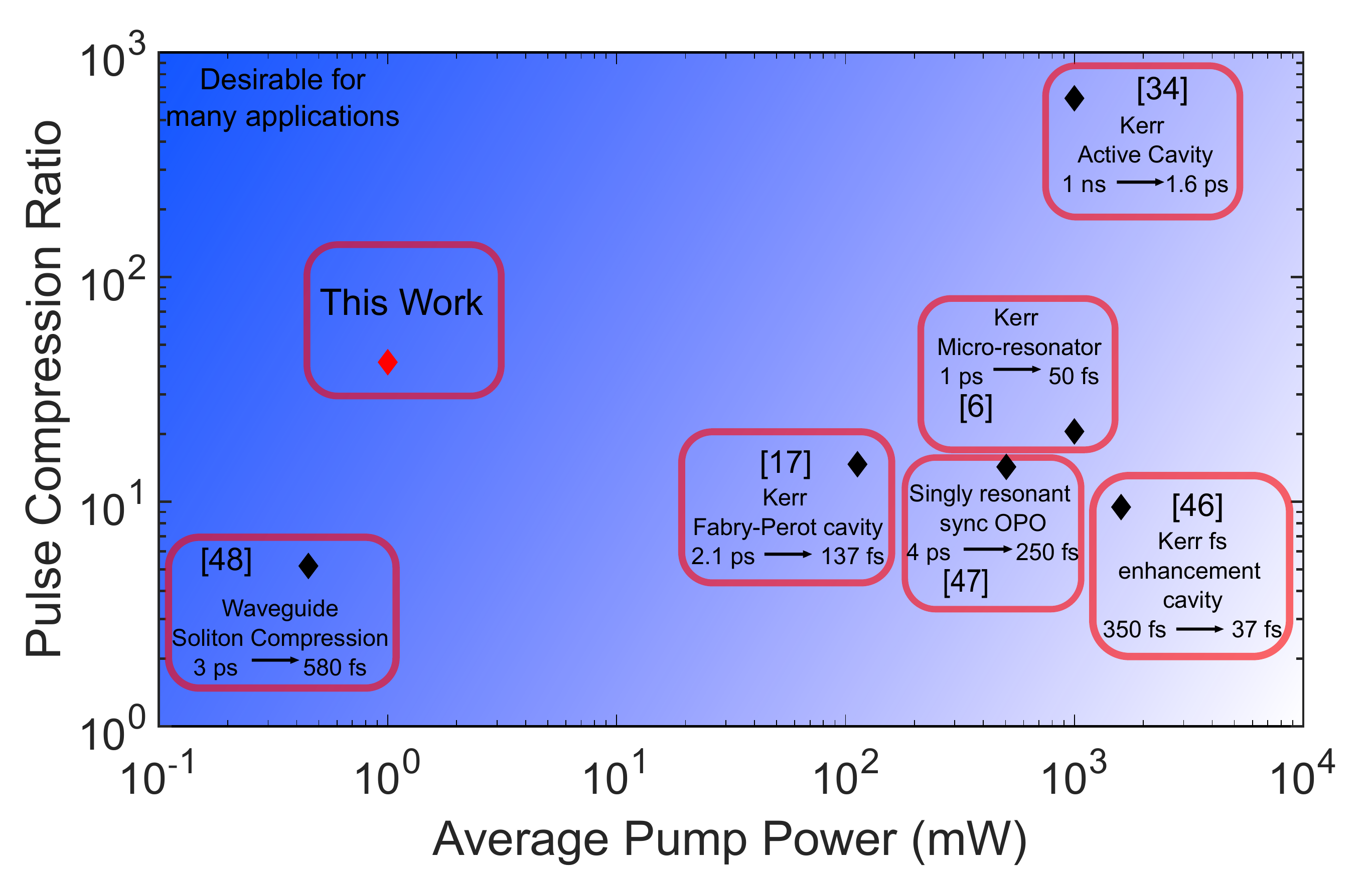}
\justifying
\caption{\label{fig: Fig5}  \textbf{Comparison of existing approaches of quadratic and cubic nonlinearity mediated pulse compression.} The $\blacklozenge $ refers to the data-points in the space of amount of pulse-compression attained against the average pump power used as obtained from the corresponding references. A desirable operating condition is the top-left corner of the plot. Here only pulse driven systems have been considered and CW-driven soliton generation process has not been included in this comparison.    }
\end{figure} 

\section{Discussion}
 
Figure 6 presents an overview of the existing approaches of pulse compression in the parameter space of compression factor and average pump power utilized. Prevailing popular approaches include, but are not limited to, pulse pumped Kerr solitons \cite{obrzud2017temporal, obrzud2019microphotonic}
, Kerr solitons in enhancement cavities \cite{lilienfein2019temporal}, Kerr solitons in active cavities \cite{englebert2021temporal}, singly resonant synchronously pumped OPOs \cite{lefort1999generation}, and soliton compression in waveguides \cite{colman2010temporal}. It is highly desirable to attain large compression factors with low pump powers (top-left corner of the landscape in Fig. 6). In our work, we have achieved a pulse-compression factor of $\sim 42$ which compressed a $\sim$ 13.2 ps flat-top pulse at 775 nm, to $\sim$ 316 fs at 1550 nm corresponding to a 3 dB spectral bandwidth of 8.3 nm (see Supplementary Section 7). The pump average power was close to 1 mW, which amounts to a meager 4 picojoule of pulse energy. This presents our work as a significant advancement over the existing approaches and elucidates the opportunities associated with soliton generation in quadratic nonlinear resonators. The experimental conversion efficiency is estimated to be near 10$\%$ (after considering the effect of insertion losses in the interfaces of the setup), which is consistent with our numerical simulation (see Supplemental Section 10). This corresponds to $\sim$4 times larger peak power of the signal compared to the pump. Additionally, extension of the demonstrated concept to integrated nano-photonic platforms will allow ultra-low power operation with soliton formation possible at several femtojoules of pump pulse energy \cite{ledezma2021intense,Lu:21, jankowski2020ultrabroadband}.\\

Various modes of operation of degenerate OPOs (DOPO) provide a rich landscape that can cater to the diverse requirements in ultra-short pulse sources in different wavelength ranges \cite{ru2017self, jankowski2018temporal}. The walk-off induced soliton represents the giant pulse-compression regime of operation of DOPO. In this solitonic regime, the pulse width scales with GDD, and good conversion efficiencies can be achieved on account of low finesse cavity operation. However, due to the gain-clipping dominated box-pulse scaling behavior, the bandwidth decreases with increasing pump power, which prevents it to achieve even higher conversion efficiencies. The simulton regime represents another solitonic mode of operation of DOPOs \cite{jankowski2018temporal}. In this regime, GVD free scaling of pulse width is observed, and thanks to the favourable trend of increasing bandwidth with pump power, higher conversion efficiency can be attained. However, the simulton regime requires that the detuning induced timing mismatch and gain-clipping window are of comparable timescale along with some minimum third order dispersion for its existence. This leads to a trade-off in terms of the attainable pulse compression and pump power(threshold requirement) (see Supplementary Section 11). Proper dispersion engineering of an OPO can lead to the formation of simultons in a dispersion regime where large pulse compression is expected to be accompanied by high ( $> 50\%$) conversion efficiencies, (see Supplementary Section 11), which will be the subject of future work. \\

In summary, we have demonstrated the formation of walk-off induced dissipative solitons in an optical parametric oscillator. We showed a significant pulse-compression factor attainable with a picojoule-level pump, thanks to the strong quadratic nonlinearity. We presented design guidelines for achieving efficient quadratic soliton formation in a properly dispersion engineered thin-film integrated devices that also allows tight optical confinement \cite{wang2018nanophotonic, jankowski2020ultrabroadband}. This holds promise for the generation of few-cycle pulses starting from easily accessible picosecond pump pulses, thereby paving the way for the realization of turn-key frequency comb sources \cite{stern2018battery,shen2020integrated}. In parallel to the ongoing efforts on pure cubic and combined quadratic and cubic nonlinearities for soliton formation \cite{englebert2021parametrically, o2020widely, bruch2020pockels}, our results highlight a route to realize solitons with pure quadratic nonlinearity with potential practical advantages in terms of conversion efficiency, wavelength diversity, and cavity finesse requirements.

\section{Methods}
\subsection{Experimental Setup}
 The simplified experimental schematic is shown in Fig 1.a, a detailed version of which is presented as Fig S.1 (Supplemental Section 2). The OPO pump is derived from the mode-locked laser through second harmonic generation (SHG) in a 40 mm long bulk quasi-phase matched periodically poled lithium niobate (PPLN) crystal. The pump is centered around 775 nm, a value which can be thermally tuned, and is $\sim$ 13 ps in duration. This provides approximately the optimum pump length for soliton formation, since it satisfies $T_{p}=Lu$. The main cavity is composed of a PPLN waveguide (reverse proton exchange, 40 mm long) \cite{langrock2007fiber} with fiber coupled output ports, fiber phase shifter, free-space section (to adjust the pump repetition rate to be multiple of the free spectral range of the cavity.), additional fiber segment to engineer the cavity dispersion (combination of polarization maintaining single mode providing anomalous dispersion and dispersion compensating fibers providing normal dispersion), and a beam splitter which provides the output coupling. The cavity length is stabilized using the Pound-Drever-Hall (PDH) locking scheme. It is to be noted that the PDH locking scheme employed in the experiment is used to prevent the slow-drift only and doesn't play a role in the timing balance/ synchronization process. The temporal characterizations of the pulses are obtained using the non-collinear sum-frequency generation cross-correlation technique in a BBO crystal. Additional details pertaining to the experimental setup/methods are provided in the supplementary information (Supplemental Section 2).   \
\subsection{System Modeling}
 The quadratic nonlinear interaction inside the PPLN waveguide (length $L$) is governed by: 
\begin{subequations}
\label{eq:whole2}
\begin{equation}
\frac{\partial a}{\partial z} =\left[ - \frac{\alpha^{(a)}}{2}  -\textrm{i}\frac{\beta_{2}^{(a)}}{2!}\frac{\partial^{2}}{\partial t^{2}}+ \ldots\right]a +\epsilon a^{*}b
\label{subeq:3}
\end{equation}
\begin{equation}
\frac{\partial b}{\partial z}=\left[- \frac{\alpha^{(b)}}{2} - u\frac{\partial}{\partial t} -\textrm{i}\frac{\beta_{2}^{(b)}}{2!}\frac{\partial^{2}}{\partial t^{2}} + \ldots \right]b -\frac{\epsilon a^{2}}{2}
\label{subeq:4}
\end{equation}
\end{subequations}

 The evolution of the signal($a$) and  the pump($b$) envelopes in the slowly varying envelope approximation are dictated by (2a) and (2b) respectively \cite{hamerly2016reduced}. Here $u$ represents the walk-off parameter, $\alpha$ denotes the loss co-efficients, and the group velocity dispersion co-efficients are denoted by $\beta$.  The effective second-order nonlinear co-efficient ($\epsilon$) is related to the SHG efficiency \cite{hamerly2016reduced}. The round-trip cavity feedback (outside the periodically poled region) is given by:
 \begin{subequations}
\label{eq:whole3}
\begin{equation}
a^{(n+1)}(0,t) = {\cal F}^{-1}\left\{ G_{0}^{-\frac{1}{2}} e^{i\bar{\phi}} {\cal F} \left\{ a^{(n)}(L,t)\right\}\right\} %
\label{subeq:5}
\end{equation} 
\begin{equation}
\bar{\phi} = \Delta\phi  + \frac{l \lambda^{(a)}}{2c}(\delta\omega)  + \frac{\phi_{2}}{2!}(\delta\omega) ^{2}+ \ldots
\label{subeq:6}
\end{equation}
\end{subequations}

 Eq. (3) takes into consideration the round-trip loss which is lumped into an aggregated out-coupling loss factor $G_{0}$, the GVD ($\phi_{2}$) of the feedback path and the detuning ($\Delta\phi$) ($\Delta\phi=\pi l$, $l$ is the cavity length detuning in units of signal half-wavelengths in vacuum) of the circulating signal from the exact synchrony with respect to the periodic pump.   The round-trip number is denoted by $n$. The equations are numerically solved adopting the split-step Fourier algorithm. Further details are provided in supplementary section 3,4. We also adopted the manifold projection method to obtain semi-analytical expression of the soliton parameters. Results obtained in this approach are referred to as analytical. \\

\section{Data availability}
The data that support the plots within this paper and other findings of this study are available from the corresponding author upon reasonable request. \\

\section{Code availability}
The codes that support the findings of this study are available from the corresponding author upon reasonable request.

\nocite{*}
\bibliographystyle{unsrt} 
\bibliography{ref}

\begin{thebibliography}{10}

\bibitem{kippenberg2018dissipative}
Tobias~J Kippenberg, Alexander~L Gaeta, Michal Lipson, and Michael~L
  Gorodetsky.
\newblock Dissipative {K}err solitons in optical microresonators.
\newblock {\em Science}, 361(6402), 2018.

\bibitem{herr2014temporal}
Tobias Herr, Victor Brasch, John~D Jost, Christine~Y Wang, Nikita~M Kondratiev,
  Michael~L Gorodetsky, and Tobias~J Kippenberg.
\newblock Temporal solitons in optical microresonators.
\newblock {\em Nature Photonics}, 8(2):145--152, 2014.

\bibitem{dutt2018chip}
Avik Dutt, Chaitanya Joshi, Xingchen Ji, Jaime Cardenas, Yoshitomo Okawachi,
  Kevin Luke, Alexander~L Gaeta, and Michal Lipson.
\newblock On-chip dual-comb source for spectroscopy.
\newblock {\em Science Advances}, 4(3):e1701858, 2018.

\bibitem{muraviev2018massively}
AV~Muraviev, VO~Smolski, ZE~Loparo, and KL~Vodopyanov.
\newblock Massively parallel sensing of trace molecules and their isotopologues
  with broadband subharmonic mid-infrared frequency combs.
\newblock {\em Nature Photonics}, 12(4):209--214, 2018.

\bibitem{suh2019searching}
Myoung-Gyun Suh, Xu~Yi, Yu-Hung Lai, S~Leifer, Ivan~S Grudinin, G~Vasisht,
  Emily~C Martin, Michael~P Fitzgerald, G~Doppmann, J~Wang, et~al.
\newblock Searching for exoplanets using a microresonator astrocomb.
\newblock {\em Nature photonics}, 13(1):25--30, 2019.

\bibitem{obrzud2019microphotonic}
Ewelina Obrzud, Monica Rainer, Avet Harutyunyan, Miles~H Anderson, Junqiu Liu,
  Michael Geiselmann, Bruno Chazelas, Stefan Kundermann, Steve Lecomte, Massimo
  Cecconi, et~al.
\newblock A microphotonic astrocomb.
\newblock {\em Nature Photonics}, 13(1):31--35, 2019.

\bibitem{stern2020direct}
Liron Stern, Jordan~R Stone, Songbai Kang, Daniel~C Cole, Myoung-Gyun Suh,
  Connor Fredrick, Zachary Newman, Kerry Vahala, John Kitching, Scott~A
  Diddams, et~al.
\newblock Direct {K}err frequency comb atomic spectroscopy and stabilization.
\newblock {\em Science Advances}, 6(9):eaax6230, 2020.

\bibitem{suh2018soliton}
Myoung-Gyun Suh and Kerry~J Vahala.
\newblock Soliton microcomb range measurement.
\newblock {\em Science}, 359(6378):884--887, 2018.

\bibitem{trocha2018ultrafast}
Philipp Trocha, M~Karpov, D~Ganin, Martin~HP Pfeiffer, Arne Kordts, S~Wolf,
  J~Krockenberger, Pablo Marin-Palomo, Claudius Weimann, Sebastian Randel,
  et~al.
\newblock Ultrafast optical ranging using microresonator soliton frequency
  combs.
\newblock {\em Science}, 359(6378):887--891, 2018.

\bibitem{ji2019chip}
Xingchen Ji, Xinwen Yao, Alexander Klenner, Yu~Gan, Alexander~L Gaeta,
  Christine~P Hendon, and Michal Lipson.
\newblock Chip-based frequency comb sources for optical coherence tomography.
\newblock {\em Optics Express}, 27(14):19896--19905, 2019.

\bibitem{marin2017microresonator}
Pablo Marin-Palomo, Juned~N Kemal, Maxim Karpov, Arne Kordts, Joerg Pfeifle,
  Martin~HP Pfeiffer, Philipp Trocha, Stefan Wolf, Victor Brasch, Miles~H
  Anderson, et~al.
\newblock Microresonator-based solitons for massively parallel coherent optical
  communications.
\newblock {\em Nature}, 546(7657):274--279, 2017.

\bibitem{xu202111}
Xingyuan Xu, Mengxi Tan, Bill Corcoran, Jiayang Wu, Andreas Boes, Thach~G
  Nguyen, Sai~T Chu, Brent~E Little, Damien~G Hicks, Roberto Morandotti, et~al.
\newblock 11 {TOPS} photonic convolutional accelerator for optical neural
  networks.
\newblock {\em Nature}, 589(7840):44--51, 2021.

\bibitem{Lu:21}
Juanjuan Lu, Ayed~Al Sayem, Zheng Gong, Joshua~B. Surya, Chang-Ling Zou, and
  Hong~X. Tang.
\newblock Ultralow-threshold thin-film lithium niobate optical parametric
  oscillator.
\newblock {\em Optica}, 8(4):539--544, Apr 2021.

\bibitem{xue2019super}
Xiaoxiao Xue, Xiaoping Zheng, and Bingkun Zhou.
\newblock Super-efficient temporal solitons in mutually coupled optical
  cavities.
\newblock {\em Nature Photonics}, 13(9):616--622, 2019.

\bibitem{xue2017microresonator}
Xiaoxiao Xue, Pei-Hsun Wang, Yi~Xuan, Minghao Qi, and Andrew~M Weiner.
\newblock Microresonator {K}err frequency combs with high conversion
  efficiency.
\newblock {\em Laser \& Photonics Reviews}, 11(1):1600276, 2017.

\bibitem{bao2014nonlinear}
Changjing Bao, Lin Zhang, Andrey Matsko, Yan Yan, Zhe Zhao, Guodong Xie,
  Anuradha~M Agarwal, Lionel~C Kimerling, Jurgen Michel, Lute Maleki, et~al.
\newblock Nonlinear conversion efficiency in {K}err frequency comb generation.
\newblock {\em Optics Letters}, 39(21):6126--6129, 2014.

\bibitem{obrzud2017temporal}
Ewelina Obrzud, Steve Lecomte, and Tobias Herr.
\newblock Temporal solitons in microresonators driven by optical pulses.
\newblock {\em Nature Photonics}, 11(9):600, 2017.

\bibitem{bao2019laser}
Hualong Bao, Andrew Cooper, Maxwell Rowley, Luigi Di~Lauro, Juan
  Sebastian~Totero Gongora, Sai~T Chu, Brent~E Little, Gian-Luca Oppo, Roberto
  Morandotti, David~J Moss, et~al.
\newblock Laser cavity-soliton microcombs.
\newblock {\em Nature Photonics}, 13(6):384--389, 2019.

\bibitem{stern2018battery}
Brian Stern, Xingchen Ji, Yoshitomo Okawachi, Alexander~L Gaeta, and Michal
  Lipson.
\newblock Battery-operated integrated frequency comb generator.
\newblock {\em Nature}, 562(7727):401--405, 2018.

\bibitem{shen2020integrated}
Boqiang Shen, Lin Chang, Junqiu Liu, Heming Wang, Qi-Fan Yang, Chao Xiang,
  Rui~Ning Wang, Jijun He, Tianyi Liu, Weiqiang Xie, et~al.
\newblock Integrated turnkey soliton microcombs.
\newblock {\em Nature}, 582(7812):365--369, 2020.

\bibitem{moille2021ultra}
Gregory Moille, Edgar~F Perez, Ashutosh Rao, Xiyuan Lu, Yanne Chembo, and
  Kartik Srinivasan.
\newblock Ultra-{B}roadband {S}oliton {M}icrocomb {T}hrough {S}ynthetic
  {D}ispersion.
\newblock {\em arXiv preprint arXiv:2102.00301}, 2021.

\bibitem{bruch2020pockels}
Alexander~W Bruch, Xianwen Liu, Zheng Gong, Joshua~B Surya, Ming Li, Chang-Ling
  Zou, Hong Tang, et~al.
\newblock Pockels {S}oliton {M}icrocomb.
\newblock {\em Nature Photonics}, 15(1):53--58, 2021.

\bibitem{mosca2016direct}
Simona Mosca, Iolanda Ricciardi, Maria Parisi, Pasquale Maddaloni, Luigi
  Santamaria, Paolo De~Natale, and Maurizio De~Rosa.
\newblock Direct generation of optical frequency combs in $\chi$ (2) nonlinear
  cavities.
\newblock {\em Nanophotonics}, 5(2):316--331, 2016.

\bibitem{mosca2018modulation}
Simona Mosca, Melissa Parisi, Iolanda Ricciardi, Fran{\c{c}}ois Leo, Tobias
  Hansson, Miro Erkintalo, Pasquale Maddaloni, Paolo De~Natale, Stefan Wabnitz,
  and Maurizio De~Rosa.
\newblock Modulation instability induced frequency comb generation in a
  continuously pumped optical parametric oscillator.
\newblock {\em Physical Review Letters}, 121(9):093903, 2018.

\bibitem{parra2019frequency}
Pedro Parra-Rivas, Lendert Gelens, Tobias Hansson, Stefan Wabnitz, and
  Fran{\c{c}}ois Leo.
\newblock Frequency comb generation through the locking of domain walls in
  doubly resonant dispersive optical parametric oscillators.
\newblock {\em Optics Letters}, 44(8):2004--2007, 2019.

\bibitem{leo2016walk}
Fran{\c{c}}ois Leo, Tobias Hansson, Iolanda Ricciardi, Maurizio De~Rosa,
  St{\'e}phane Coen, Stefan Wabnitz, and Miro Erkintalo.
\newblock Walk-off-induced modulation instability, temporal pattern formation,
  and frequency comb generation in cavity-enhanced second-harmonic generation.
\newblock {\em Physical Review Letters}, 116(3):033901, 2016.

\bibitem{nie2021deterministic}
Mingming Nie, Yijun Xie, and Shu-Wei Huang.
\newblock Deterministic generation of parametrically driven dissipative {K}err
  soliton.
\newblock {\em Nanophotonics}, 10(6):1691--1699, 2021.

\bibitem{nie2020quadratic}
Mingming Nie and Shu-Wei Huang.
\newblock Quadratic soliton mode-locked degenerate optical parametric
  oscillator.
\newblock {\em Optics Letters}, 45(8):2311--2314, 2020.

\bibitem{leindecker2012octave}
Nick Leindecker, Alireza Marandi, Robert~L Byer, Konstantin~L Vodopyanov, Jie
  Jiang, Ingmar Hartl, Martin Fermann, and Peter~G Schunemann.
\newblock Octave-spanning ultrafast {OPO} with 2.6-6.1 $\mu$m instantaneous
  bandwidth pumped by femtosecond {T}m-fiber laser.
\newblock {\em Optics Express}, 20(7):7046--7053, 2012.

\bibitem{jankowski2018temporal}
Marc Jankowski, Alireza Marandi, CR~Phillips, Ryan Hamerly, Kirk~A Ingold,
  Robert~L Byer, and MM~Fejer.
\newblock Temporal simultons in optical parametric oscillators.
\newblock {\em Physical Review Letters}, 120(5):053904, 2018.

\bibitem{cole2018kerr}
Daniel~C Cole, Jordan~R Stone, Miro Erkintalo, Ki~Youl Yang, Xu~Yi, Kerry~J
  Vahala, and Scott~B Papp.
\newblock Kerr-microresonator solitons from a chirped background.
\newblock {\em Optica}, 5(10):1304--1310, 2018.

\bibitem{jang2015temporal}
Jae~K Jang, Miro Erkintalo, Stephane Coen, and Stuart~G Murdoch.
\newblock Temporal tweezing of light through the trapping and manipulation of
  temporal cavity solitons.
\newblock {\em Nature Communications}, 6(1):1--7, 2015.

\bibitem{erkintalo2021phase}
Miro Erkintalo, Stuart~G Murdoch, and St{\'e}phane Coen.
\newblock Phase and intensity control of dissipative {K}err cavity solitons.
\newblock {\em Journal of the Royal Society of New Zealand}, pages 1--19, 2021.

\bibitem{englebert2021temporal}
Nicolas Englebert, Carlos~Mas Arab{\'\i}, Pedro Parra-Rivas, Simon-Pierre
  Gorza, and Fran{\c{c}}ois Leo.
\newblock Temporal solitons in a coherently driven active resonator.
\newblock {\em Nature Photonics}, pages 1--6, 2021.

\bibitem{hamerly2016reduced}
Ryan Hamerly, Alireza Marandi, Marc Jankowski, Martin~M Fejer, Yoshihisa
  Yamamoto, and Hideo Mabuchi.
\newblock Reduced models and design principles for half-harmonic generation in
  synchronously pumped optical parametric oscillators.
\newblock {\em Physical Review A}, 94(6):063809, 2016.

\bibitem{langrock2007fiber}
Carsten Langrock and MM~Fejer.
\newblock Fiber-feedback continuous-wave and synchronously-pumped
  singly-resonant ring optical parametric oscillators using
  reverse-proton-exchanged periodically-poled lithium niobate waveguides.
\newblock {\em Optics Letters}, 32(15):2263--2265, 2007.

\bibitem{marandi2015guided}
Alireza Marandi, Carsten Langrock, Martin~M Fejer, and Robert~L Byer.
\newblock Guided-wave half-harmonic generation of frequency combs with~ 75-fold
  spectral broadening.
\newblock In {\em Nonlinear Optics}, pages NM1A--2. Optical Society of America,
  2015.

\bibitem{akhmanov1968nonstationary}
S~Akhmanov, A~Chirkin, K~Drabovich, A~Kovrigin, R~Khokhlov, and A~Sukhorukov.
\newblock Nonstationary nonlinear optical effects and ultrashort light pulse
  formation.
\newblock {\em IEEE Journal of Quantum Electronics}, 4(10):598--605, 1968.

\bibitem{khaydarov1995pulse}
John~DV Khaydarov, James~H Andrews, and Kenneth~D Singer.
\newblock Pulse-compression mechanism in a synchronously pumped optical
  parametric oscillator.
\newblock {\em JOSA B}, 12(11):2199--2208, 1995.

\bibitem{ledezma2021intense}
Luis Ledezma, Ryoto Sekine, Qiushi Guo, Rajveer Nehra, Saman Jahani, and
  Alireza Marandi.
\newblock Intense optical parametric amplification in dispersion engineered
  nanophotonic lithium niobate waveguides.
\newblock {\em arXiv preprint arXiv:2104.08262}, 2021.

\bibitem{roy2021spectral}
Arkadev Roy, Saman Jahani, Carsten Langrock, Martin Fejer, and Alireza Marandi.
\newblock Spectral phase transitions in optical parametric oscillators.
\newblock {\em Nature Communications}, 12(1):1--9, 2021.

\bibitem{trillo1996bright}
Stefano Trillo.
\newblock Bright and dark simultons in second-harmonic generation.
\newblock {\em Optics Letters}, 21(15):1111--1113, 1996.

\bibitem{jankowski2020ultrabroadband}
Marc Jankowski, Carsten Langrock, Boris Desiatov, Alireza Marandi, Cheng Wang,
  Mian Zhang, Christopher~R Phillips, Marko Lon{\v{c}}ar, and MM~Fejer.
\newblock Ultrabroadband nonlinear optics in nanophotonic periodically poled
  lithium niobate waveguides.
\newblock {\em Optica}, 7(1):40--46, 2020.

\bibitem{jankowski2021efficient}
Marc Jankowski, Nayara Jornod, Carsten Langrock, Boris Desiatov, Alireza
  Marandi, Marko Lon{\v{c}}ar, and Martin~M Fejer.
\newblock Efficient {O}ctave-{S}panning {P}arametric {D}own-{C}onversion at the
  {P}icojoule level.
\newblock {\em arXiv preprint arXiv:2104.07928}, 2021.

\bibitem{dietrich2020higher}
Christian~M Dietrich, Ihar Babushkin, Jos{\'e} R~Cardoso de~Andrade, Han Rao,
  Ayhan Demircan, and Uwe Morgner.
\newblock Higher-order dispersion and the spectral behavior in a doubly
  resonant optical parametric oscillator.
\newblock {\em Optics Letters}, 45(20):5644--5647, 2020.

\bibitem{lilienfein2019temporal}
Nikolai Lilienfein, Christina Hofer, Maximilian H{\"o}gner, Tobias Saule,
  Michael Trubetskov, Volodymyr Pervak, Ernst Fill, Claudius Riek, Alfred
  Leitenstorfer, J~Limpert, et~al.
\newblock Temporal solitons in free-space femtosecond enhancement cavities.
\newblock {\em Nature Photonics}, 13(3):214--218, 2019.

\bibitem{lefort1999generation}
L~Lefort, K~Puech, SD~Butterworth, YP~Svirko, and DC~Hanna.
\newblock Generation of femtosecond pulses from order-of-magnitude pulse
  compression in a synchronously pumped optical parametric oscillator based on
  periodically poled lithium niobate.
\newblock {\em Optics Letters}, 24(1):28--30, 1999.

\bibitem{colman2010temporal}
Pierre Colman, Chad Husko, Sylvain Combri{\'e}, Isabelle Sagnes, Chee~Wei Wong,
  and Alfredo De~Rossi.
\newblock Temporal solitons and pulse compression in photonic crystal
  waveguides.
\newblock {\em Nature Photonics}, 4(12):862, 2010.

\bibitem{ru2017self}
Qitian Ru, Zachary~E Loparo, XiaoSheng Zhang, Sean Crystal, Subith Vasu,
  Peter~G Schunemann, and Konstantin~L Vodopyanov.
\newblock Self-referenced octave-wide subharmonic {G}a{P} optical parametric
  oscillator centered at 3 $\mu$m and pumped by an {E}r-fiber laser.
\newblock {\em Optics Letters}, 42(22):4756--4759, 2017.

\bibitem{wang2018nanophotonic}
Cheng Wang, Mian Zhang, Brian Stern, Michal Lipson, and Marko Lon{\v{c}}ar.
\newblock Nanophotonic lithium niobate electro-optic modulators.
\newblock {\em Optics express}, 26(2):1547--1555, 2018.

\bibitem{englebert2021parametrically}
Nicolas Englebert, Francesco De~Lucia, Pedro Parra-Rivas, Carlos~Mas
  Arab{\'\i}, Pier-John Sazio, Simon-Pierre Gorza, and Fran{\c{c}}ois Leo.
\newblock Parametrically driven kerr cavity solitons.
\newblock {\em arXiv preprint arXiv:2101.07784}, 2021.

\bibitem{o2020widely}
Callum~F O’Donnell, S~Chaitanya Kumar, Therese Paoletta, and Majid
  Ebrahim-Zadeh.
\newblock Widely tunable femtosecond soliton generation in a fiber-feedback
  optical parametric oscillator.
\newblock {\em Optica}, 7(5):426--433, 2020.

\end{thebibliography}

\section{Acknowledgments}
 The authors gratefully acknowledge support from AFOSR award FA9550-20-1-0040, NSF Grant No. 1846273, and NASA. The authors wish to thank NTT Research for their financial and technical support. The authors thank Robert Gray for his valuable inputs.\\

\section{Author Contribution}
 A.R. performed the experiments with help from R.N.   A.R., and L.L. developed the theory and performed the numerical simulations. C.L. fabricated the PPLN waveguide used in the experiment with supervision of M.F. All authors contributed to the analysis of the results. A.R. and A.M. wrote the manuscript with input from all authors. A.M. supervised the project. \\

\section{Competing Interests}
The authors declare no competing interests. \\

\end{document}